\documentclass[prd,aps,showpacs,tightenlines,nofootinbib,preprint]{revtex4}
\usepackage{graphicx}
\usepackage{amsfonts}
\usepackage{amssymb}
\usepackage{latexsym}

\newcommand{\CO}{{\cal O}}

\newcommand{\bear}{\begin{array}}  \newcommand{\eear}{\end{array}}
\newcommand{\bea}{\begin{eqnarray}}  \newcommand{\eea}{\end{eqnarray}}
\newcommand{\beq}{\begin{equation}}  \newcommand{\eeq}{\end{equation}}
\newcommand{\bef}{\begin{figure}}  \newcommand{\eef}{\end{figure}}
\newcommand{\bec}{\begin{center}}  \newcommand{\eec}{\end{center}}
\newcommand{\non}{\nonumber}  
\newcommand{\lmk}{\left(}  \newcommand{\rmk}{\right)}
\newcommand{\lkk}{\left[}  \newcommand{\rkk}{\right]}

\newcommand{\bib}{\bibitem} 
\newcommand{\la}{\left\langle} \newcommand{\ra}{\right\rangle}

\newcommand{\ns}{n_s}
\newcommand{\Mpc}{{\rm Mpc}}
\newcommand{\mg}{M_G}

\newcommand{\muni}{\mu^2}
\newcommand{\kx}{c_X}
\newcommand{\kz}{c_Z}
\newcommand{\kp}{c_{\Psi}}
\newcommand{\kn}{c_N}

\newcommand{\Phimin}{\Phi_{\min}}
\newcommand{\calr}{{\cal R}}

\def\IBB#1#2#3{{\bf #1}, #2 (20#3)}

\def\IBIDD#1#2#3{{\it ibid}. {\bf #1}, #2 (20#3)}

\def\JL#1#2#3{JETP. Lett. {\bf #1}, #2 (19#3)}

\def\JHEPP#1#2#3{J. High Energy Phys. {\bf #1}, #2 (20#3)}

\def\PLBold#1#2#3{Phys. Lett. {\bf#1B}, #2 (19#3)}

\def\PRD#1#2#3{Phys. Rev. D {\bf #1}, #2 (19#3)}
\def\PRDD#1#2#3{Phys. Rev. D {\bf #1}, #2 (20#3)}

\def\PRLL#1#2#3{Phys. Rev. Lett. {\bf#1}, #2 (20#3)}

\begin{document}
\title{Chaotic hybrid new inflation in supergravity \\ with a running
spectral index}

\author{Masahide Yamaguchi} 
\affiliation{Physics Department, Brown University, Providence, Rhode Island
  02912, USA}
\author{Jun'ichi Yokoyama} 
\affiliation{Department of Earth and Space Science, Graduate School of
  Science, Osaka University, Toyonaka 560-0043, Japan}

\date{\today}


\begin{abstract}
  We propose an inflation model in supergravity, in which chaotic and
  hybrid inflation occurs successively, followed by new inflation.
  During hybrid inflation, adiabatic fluctuations with a running
  spectral index with $\ns >1$ on a large scale and $\ns <1$ on a
  smaller scale are generated, as favored by recent results of the
  first year Wilkinson Microwave Anisotropy Probe. The initial
  condition of new inflation is also set dynamically during hybrid
  inflation, and its duration and the amplitude of density
  fluctuations take appropriate values to help early star formation to
  realize early reionization.
\end{abstract}

\pacs{98.80.Cq,04.65.+e,11.27.+d \hspace{4.0cm} BROWN-HET-1371, OU-TAP-216}

\maketitle


\section{Introduction}

\label{sec:intro}

The recent results of the Wilkinson Microwave Anisotropy Probe (WMAP)
confirmed the so-called concordance model, in which the Universe is
spatially flat and has primordial adiabatic density fluctuations
\cite{Bennett:2003bz,Spergel:2003cb}. The spectrum of the density
fluctuations is almost scale invariant but the running of the spectral
index is favored from $\ns>1$ on a large scale to $\ns<1$ on a smaller
scale.\footnote{Such a running of the spectral index is not yet
  confirmed definitely. The analyses which do not give the running
  feature of the spectral index are given, for example, in
  \cite{KKMR,LL}. However, it is still important to give a concrete
  model with such a running spectral index because it is very
  difficult to realize such a feature in the context of a single-field
  inflation model.} More concretely, it is shown that $\ns=1.13\pm
0.08$ and $d\ns/d\ln k=-0.055^{+0.028}_{-0.029}$ on the scale
$k_0=0.002\,\Mpc^{-1}$ \cite{Peiris:2003ff}.

Though it is a nontrivial task to consider an inflation model with
such a running spectral index, some models have been proposed mainly
after the results of the WMAP have been released
\cite{Linde:1997sj,FLZZ,KS,KYY,HL,CST}. In the hybrid inflation model
in supergravity proposed by Linde and Riotto \cite{Linde:1997sj}, the
running of the spectral index is obtained straightforwardly due to the
contributions to the potential from both one-loop effects and
supergravity effects. However, it is shown that its variation is too
mild to explain the results of the WMAP \cite{KYY}. This is mainly
because the Yukawa coupling constant must be relatively small for
sufficient inflation. To put it another way, a sufficient variation of
the running may be obtained if another inflation follows the hybrid
inflation of Linde and Riotto. Then, Kawasaki and the present authors
considered a hybrid new inflation model in supergravity, which was
originally proposed in
\cite{Izawa:1997df,Kawasaki:1998vx,Kanazawa:1999ag}, and showed that
the results of the WMAP can be reproduced at the one-sigma level
\cite{KYY}.  In this model, during hybrid inflation, primordial
density fluctuations with such a running spectral index are generated
and also the initial value of new inflation is set dynamically. If new
inflation proceeds long enough, the scales with the desired spectral
shape of density fluctuations can be pushed to cosmologically
observable scales.

This model, however, has two drawbacks.  First of all, hybrid
inflation has a severe initial condition problem, that is, only very
limited field configurations lead to inflation even if the field is
assumed to be homogeneous \cite{hybridinit}.  Second, the density
fluctuations generated at the onset of new inflation tends to be too
large due to the uncontrollable smallness of the initial value of the
scalar field responsible for new inflation.  In the model discussed in
\cite{KYY}, if we try to fit the central values of the spectral index
and its running observed by WMAP, the scale $l_{\ast}$ corresponding
to the horizon at the onset of new inflation is larger than 100 kpc,
so that too many dark-halo-like objects would be produced and cause
cosmological problems.  Hence we had to content ourselves with
reproducing the WMAP data within one-sigma error in order to set
$l_{\ast}$ smaller than 1 kpc.

In this paper, we propose an improved model of inflation in
supergravity, which resolves the above two problems.  With regards to
the former problem associated with initial conditions, the most
natural scenario is chaotic inflation which occurs without any
fine-tuning provided that the potential does not diverge too rapidly
beyond the gravitational scale $M_{G} \simeq 2.4 \times 10^{18}$ GeV
\cite{Linde:gd}.  This requirement, however, is difficult to achieve
in supergravity and only a few successful models had been proposed in
this context \cite{GM}.  Recently, however, it was pointed out that
such a large value can be realized by introducing the
Nambu-Goldstone-like shift symmetry \cite{chaotic,double}.  We make
use of this symmetry and start with chaotic inflation.  Furthermore,
in our model, the initial condition for new inflation, which is set
dynamically during hybrid inflation, is well under control and we can
set it to satisfy various cosmological requirements with the central
values obtained by WMAP, contrary to our previous model.

\if
In our model, hybrid inflation takes place
with a chaotic initial condition, that is, hybrid inflation begins with
a value much larger than the gravitational scale $M_{G} \simeq 2.4
\times 10^{18}$ GeV. Generally speaking, it is very difficult to have a
value larger than the gravitational scale in supergravity because the
potential has the exponential growth beyond the gravitational scale if
we take a canonical K\"ahler potential. However, it was pointed out that
such a large value can be realized by introducing the Nambu-Goldstone
like shift symmetry \cite{chaotic,double}. Furthermore, in our model,
the initial value of new inflation, which is set dynamically during
hybrid inflation, becomes small enough to acquire a large e-fold number
but large enough to suppress primordial density fluctuations not to
overproduce dark halo objects on corresponding scales.
\fi

The rest of the paper is organized as follows. In the next section, we
present our model of hybrid new inflation with a chaotic initial
condition in supergravity. In Sec. \ref{sec:hybrid}, we investigate
the dynamics of hybrid inflation and show that primordial density
fluctuations with a running spectral index are generated, as suggested
by the results of the WMAP. In Sec. \ref{sec:new}, we review new
inflation in supergravity. In Sec. \ref{sec:hybridnew}, we investigate
the dynamics of hybrid new inflation in supergravity. We show that the
initial value of new inflation is set dynamically during hybrid
inflation and the amplitude of density fluctuations on the scale
corresponding to the horizon at the onset of new inflation takes an
appropriate value.  Section VI is devoted to discussion and conclusion
with particular emphasis on the fact that this model essentially
involves only one energy scale.

\section{Model}

\label{sec:model}

In this section, we give our model of chaotic hybrid new inflation in
supergravity.  In order to avoid blow up of the scalar potential such
as $\exp\lmk |S|^2/M_G^2\rmk$ for the inflaton, $S$, responsible for
chaotic and hybrid inflation, we introduce the Nambu-Goldstone-like
shift symmetry \cite{chaotic},
\beq
  S \rightarrow S + i~C M_{G},
\eeq
where  $C$ is a dimensionless real constant. As
far as the symmetry is exact, however, the field $S$ cannot have any potential.
So we need to break it softly. By introducing a spurion superfield
$\Xi$, we extend the shift symmetry such that the model is invariant
under the following transformation \cite{KYY}:
\bea
  S &\rightarrow& S + i~C M_{G}, \non \\
  \Xi  &\rightarrow& \frac{S}{S + i~C M_{G}} \Xi.
  \label{eq:shift}
\eea
Then, the combination $\Xi S$ is invariant under the shift symmetry
and the K\"ahler potential becomes a function of $S + S^{\ast}$, i.e.,
$K(S,S^{\ast}) = K(S + S^{\ast})$, which allows the imaginary part of
the scalar components of $S$ to take a value much larger than $M_{G}$.
If the spurion field acquires a vacuum expectation value, $\la \Xi \ra
\ll M_{G}$, it softly breaks the above shift symmetry.  Below we set
$M_{G}$ to unity and use the same notations for scalar fields and
corresponding superfields.

We consider the following superpotential comprised of three parts:
\bea
  W &=& W_H + W_I + W_N, \non \\
    &W_H& = X( \lambda'\Psi\overline\Psi - \mu'^2 \Pi^2 ) 
          + g' \Xi S \Psi\overline\Psi + \nu' Z \Xi^2 S^2  , \non \\
    &W_I& = u' \Pi \Sigma Z \Phi,
             \non \\ 
    &W_N& = v'^2 \Pi^4 \Phi - \frac{h'}{n+1}\Phi^{n+1}.  \label{dash}
\eea
Here $W_H$ induces chaotic and hybrid inflation, $W_I$ determines the
initial value of the new inflation, and $W_N$ is a part associated
with new inflation.  In the above superpotential, $\Psi$ and
$\overline\Psi$ are a conjugate pair of superfields, which transform
nontrivially under some gauge group $G$, while the other superfields
are gauge singlet. In order to obtain such a superpotential, we
introduce the U$(1)_{R}$ symmetry and the $Z_2$ symmetry, which are
also broken softly by introducing the spurion fields $\Pi$ and
$\Sigma$, in addition to the shift symmetry. $\lambda'$, $\mu'$, $g'$,
$\nu'$, $u'$, $v'$, and $h'$ are coupling constants.  The charge
assignments for various superfields are shown in Table I.  Note that
all softly broken symmetries are restored when all the spurion fields
have vanishing expectation values, $\la \Xi\ra=\la \Pi\ra=\la
\Sigma\ra=0$.  If $\Xi$ acquires a nonvanishing expectation value, it
breaks not only the shift symmetry but also the U$(1)_{R}$ symmetry
and the $Z_2$ symmetry, so that we may expect that the magnitudes of
the breaking of the three symmetries are mutually related.  We will
show later that even if we take a simple view that all the symmetry
breaking scales are identical, $\la \Xi\ra=\la \Pi\ra=\la
\Sigma\ra=\CO(10^{-2})$, our model can reproduce various cosmological
observations quite well with all the other model parameters being
within a natural range of $10^{-3}\sim 1$.  Then, by inserting the
vacuum expectation values of the spurion fields and neglecting higher
order terms, we have the superpotential\footnote{Terms proportional to
  $\Xi S \Pi^2$ and $ZX^2$ can appear but here we have omitted them
  because they do not have significant effects on the dynamics, as can
  be easily shown.}
\bea
  W &=& W_H + W_I + W_N, \non \\
    &W_H& = X( \lambda\Psi\overline\Psi - \mu^2 ) 
          + g S \Psi\overline\Psi + \nu Z S^2  , \non \\
    &W_I& = u Z \Phi,
             \non \\ 
    &W_N& = v^2 \Phi - \frac{h}{n+1}\Phi^{n+1}.
    \label{eq:totalsuper}
\eea
Here $\lambda\equiv\lambda'=\CO(10^{-1})$, 
$\mu\equiv\mu'\la\Pi\ra=\CO(10^{-3})$, 
$g\equiv g'\la\Xi\ra=\CO(10^{-2})$,  
$\nu \equiv \nu' \la\Xi\ra^2 = \CO(10^{-6})$,
 $u \equiv u'\la\Pi\ra\la\Sigma\ra 
= \CO(10^{-7})$,
$v \equiv v'\la\Pi\ra^2 = \CO(10^{-6})$, and $h\equiv h'=\CO(10^{-1})$.

On the other hand, the K\"ahler potential is given by\footnote{Terms
  associated with the breaking of the U$(1)_R$ symmetry can also
  appear but here we have omitted them because they do not have
  significant effects on the dynamics.}
\bea
  K &=& K_H + K_N + \cdots, \non \\
    &K_H& = \frac12 (S + S^{\ast})^2 
      + |X|^2 - \frac{\kx}{4}|X|^4
      + |Z|^2 - \frac{\kz}{4}|Z|^4
      + |\Psi|^2 - \frac{\kp}{4}|\Psi|^4
      + |\overline\Psi|^2 - \frac{\kp}{4}|\overline\Psi|^4,  \non \\
    &K_N& = |\Phi|^2 + \frac{\kn}{4}|\Phi|^4,
    \label{eq:totalK}
\eea
where $\kx, \kz, \kp$, and $\kn$ are constants of the order of unity.

The potential of scalar components of the superfields $z_{i}$ in
supergravity is given by
\beq
  V = e^{K} \left\{ \left(
      \frac{\partial^2K}{\partial z_{i}\partial z_{j}^{*}}
    \right)^{-1}D_{z_{i}}W D_{z_{j}^{*}}W^{*}
    - 3 |W|^{2}\right\} + V_{D},
  \label{eq:potential}
\eeq
with 
\beq
  D_{z_i}W = \frac{\partial W}{\partial z_{i}} 
    + \frac{\partial K}{\partial z_{i}}W.
  \label{eq:DW}
\eeq
Here $V_{D}$ represents the D-term contribution given by
\beq
  V_{D} = \frac{e^2}{2} (|\Psi|^2 - |\overline{\Psi}|^2)^2,
\eeq 
in which we assume for simplicity that the gauge group $G$ is U$(1)$
and $e$ is the gauge coupling constant. Then, the D-term contribution
disappears for the direction $|\Psi| = |\overline{\Psi}|$.

\section{Hybrid inflation following chaotic inflation in supergravity}

\label{sec:hybrid}

In this section, we investigate the dynamics of hybrid inflation
starting with a chaotic scenario.  So, we concentrate only on $W_H$
and $K_H$ in the whole superpotential $W$ and K\"ahler potential $K$.
This treatment is justified because the energy scale of new inflation
turns out to be much smaller than chaotic and hybrid inflation as will
be seen later.

First of all, we decompose the scalar field $S$ into real and imaginary
components,
\beq
  S = \frac{1}{\sqrt{2}} (\varphi + i \sigma).
\eeq
Due to the Nambu-Goldstone-like shift symmetry, $\sigma$ does not
receive the exponential growth of the potential so that it can take a
value much larger than unity under the chaotic initial condition
\cite{Linde:gd}. Thus the potential is dominated by the quartic term
of $\sigma$, that is,
\beq
  V \simeq \frac14 \nu^2 \sigma^4,
\eeq
which leads to chaotic inflation. As $\sigma$ rolls down along the
potential, the false vacuum energy $\mu^4$ becomes dominant and the
usual hybrid inflation takes place successively.

During inflation, the mass terms of the other fields are given by
\bea
  V &\supset& \lmk \frac{\nu^2}{4}\sigma^4 + \mu^4 \rmk \varphi^2
            + \lmk \frac{\nu^2}{4}\sigma^4 + \kx \mu^4 \rmk |X|^2
            \non \\
         && + \lmk \kz \frac{\nu^2}{4}\sigma^4 + \mu^4 
                  + 2 \nu^2 \sigma^2 \rmk |Z|^2 
            + \lmk \frac{\nu^2}{4}\sigma^4 + \mu^4 
                 + \frac{g^2}{2} \sigma^2 \rmk 
            ( |\Psi|^2 + |\overline{\Psi}|^2 ) \non \\
         && - \lambda \mu^2 ( \Psi\overline{\Psi} 
              + \Psi^{\ast}\overline{\Psi}^{\ast} )
          - \frac{\nu}{2} \sigma^2 \mu^2 ( X Z^{\ast} + X^{\ast} Z ) 
          \non \\
          &=&  \lmk \frac{\nu^2}{4}\sigma^4 + \mu^4 \rmk \varphi^2
            + \lmk \frac{\nu^2}{4}\sigma^4 + \kx \mu^4 \rmk 
              \left| X - \frac{\frac{\nu}{2}\mu^2\sigma^2}
                         {\frac{\nu^2}{4}\sigma^4 + \kx \mu^4} Z \right|^2
            \non \\
         && + \frac{|Z|^2}{\frac{\nu^2}{4}\sigma^4 + \kx \mu^4}
              \lmk \frac{\kz}{16}\nu^4\sigma^8 
                   +\frac14\kx\kz\mu^4\nu^2\sigma^4 + \kx\mu^8 \rmk
            \non \\
         && + M_{-}^2 |\Psi_{+}|^2 + M_{+}^2 |\Psi_{-}|^2, 
\eea
where
\bea
  M_{\pm}^2 &=& \pm \lambda\mu^2 + \frac{\nu^2}{4}\sigma^4 
               + \mu^4 + \frac12 g^2 \sigma^2, \non \\
    \Psi_{\pm} &=& \frac{1}{\sqrt{2}}(\Psi \pm \overline{\Psi}^{\ast}). 
\eea
Then, during hybrid inflation, the fields $\varphi$, $X$, $Z$, $\Psi$, 
and $\overline{\Psi}$ rapidly go to zero for $\kx,~ \kz > 0$. $M_{-}^2$
becomes negative for $\sigma \simeq \pm \sigma_c$ with $\sigma_c \equiv
\sqrt{2\lambda}\mu/g$ because $\nu^2\sigma_c^4/4,~ \mu^4 \ll
\lambda\mu^2$. Therefore, at $\sigma \simeq \pm \sigma_c$, the phase
transition takes place and hybrid inflation ends.

During hybrid inflation, the effective potential for $\sigma$ is given
by
\beq
  V(\sigma) = \mu^4 + \frac{\nu^2}{4}\sigma^4 + V_{1L}. \label{spot}
\eeq
Here $V_{1L}$ is the one-loop correction given by \cite{Dvali:ms}
\beq
 V_{1L}=\frac{g^4}{128\pi^2}\lkk (g^2\sigma^2+2\lambda\muni)^2
 \ln\frac{g^2\sigma^2+2\lambda\muni}{\Lambda^2}
+(g^2\sigma^2-2\lambda\muni)^2
 \ln\frac{g^2\sigma^2-2\lambda\muni}{\Lambda^2}
-2g^4\sigma^4\ln\frac{g^2\sigma^2}{\Lambda^2}\rkk,
\eeq
where $\Lambda$ is the renormalization scale.  When $\sigma \gg
\sigma_c$, it is approximated as
\beq
 V_{1L}\cong \frac{\lambda^2\mu^4}{8\pi^2}\ln\frac{\sigma}{\sigma_c}.
 \label{Lapprox}
\eeq

Comparing the derivative
of the second term and that of the last term in the effective potential
(\ref{spot}), we find that the dynamics of the scalar field is dominated
by the quartic term for $\sigma > \sigma_d \equiv
\mu\sqrt{\lambda/(\sqrt{8}\pi\nu)}$ and by the radiative correction for
$\sigma < \sigma_d$. Since $\nu \sim g^2$, $\sigma_c$ and $\sigma_d$ are
comparable. Here we require $\sigma_c \lesssim \sigma_d (\ll 1)$.

The slow-roll parameters $\epsilon$, $\eta$, and $\xi$ are given by
\bea
  \epsilon &\equiv& \frac{1}{2}
         \lmk\frac{V'[\sigma]}{V[\sigma]}\rmk^2
         \cong \frac{\nu^4}{2\mu^8}\sigma^6
                \lkk 1 + \lmk \frac{\sigma_d}{\sigma} \rmk^4 \rkk^2
         = \CO(\sigma^6), \non \\
  \eta &\equiv&
         \frac{V''[\sigma]}{V[\sigma]}
         \cong \frac{\nu^2}{\mu^4}\sigma^2
                \lkk 3 - \lmk \frac{\sigma_d}{\sigma} \rmk^4 \rkk
         = \CO(\sigma^2), \non \\
  \xi  &\equiv&
         \frac{V'''[\sigma]V'[\sigma]}{V[\phi]^2}
         \cong \frac{2\nu^4}{\mu^8}\sigma^4
                \lkk 1 + \lmk \frac{\sigma_d}{\sigma} \rmk^4 \rkk
                \lkk 3 + \lmk \frac{\sigma_d}{\sigma} \rmk^4 \rkk
         = \CO(\sigma^4).
\eea
On the other hand, the amplitude of curvature perturbation in the
comoving gauge $\calr$ \cite{Bardeen} generated on the comoving scale
$r=2\pi/k$ is given by
\beq
 \calr (k)=\frac{1}{2\pi}\frac{H^2(t_k)}{|\dot{\sigma}(t_k)|},~~~
 H^2(t_k)=\frac{V[\sigma(t_k)]}{3\mg^2},
 \label{eq:fluc}
\eeq
where $t_k$ is the epoch when mode $k$ left the Hubble radius during
inflation \cite{pert}.  The spectral index and its running are given
by
\bea
  \ns-1 &=& -6\epsilon+2\eta \cong 2\eta \cong
          \frac{2\nu^2}{\mu^4}\sigma^2
                \lkk 3 - \lmk \frac{\sigma_d}{\sigma} \rmk^4 \rkk, 
          \non \\  
  \frac{d\ns}{d\ln k} &=& 16\epsilon\eta-24\epsilon^2-2\xi
          \cong - 2\xi \cong
              - \frac{4\nu^4}{\mu^8}\sigma^4
                \lkk 1 + \lmk \frac{\sigma_d}{\sigma} \rmk^4 \rkk
                \lkk 3 + \lmk \frac{\sigma_d}{\sigma} \rmk^4 \rkk.
\eea
$\sigma$ and $\sigma_d$ can be expressed by $\ns-1$ and $d\ns/d\ln k$
at $k=k_0$ as
\bea
  \sigma &=& \sqrt{\frac{\ns-1}{2(3-q)}} \frac{\mu^2}{\nu}, \non \\
  \sigma_d &=& q^{\frac14} \sigma, \non \\
\eea
where
\beq
  q \equiv \lmk\frac{\sigma_d}{\sigma}\rmk^4 = 
       \frac{3x+2 \pm \sqrt{24x+1}}{x-1} \label{20} 
\eeq
with
\bea
  x \equiv - \frac{1}{(\ns-1)^2}\frac{d\ns}{d\ln k}. \non 
\eea
Inserting the central values obtained by WMAPext+2dFGRS+Ly$\alpha$ on
a comoving scale, $k_0=0.002\,{\rm Mpc}^{-1}$, $\ns-1=0.13$ and
$d\ns/d\ln k=-0.055$ into the above equations, we find $\sigma \simeq
0.19 \mu^2/\nu$ and $\sigma_d \simeq 0.21 \mu^2/\nu$ with
$\sigma_d/\sigma \simeq 1.1$, which gives the relation
\beq
  \frac{\mu^2}{\lambda\nu} \simeq 2.7,  \label{mln}
\eeq
taking the lower sign in Eq. (\ref{20}). Also, the amplitude of
curvature perturbation in the comoving gauge $\calr$ at $\sigma$ is
given by
\beq
 \calr (k_0)= \frac{1}{2\sqrt{3}\pi} 
              \frac{\sigma \mu^6}{\nu^2(\sigma_d^4+\sigma^4)}
            \simeq 5.6 \nu
            = 4.7 \times 10^{-5}
\eeq
corresponding to $A=0.75$ of \cite{Peiris:2003ff}, which yields 
\bea
  \nu &\simeq& 8.4\times 10^{-6}, \non \\ 
  \frac{\mu^2}{\lambda} &\simeq& 2.2 \times 10^{-5}.
  \label{eq:constraint0}
\eea

The $e$-folds of hybrid inflation after the comoving scale with the
observed spectral shape has crossed the Hubble radius can be estimated as
\beq
 N_H = \int_{\sigma_c}^{\sigma} \frac{V(\sigma)}{V'(\sigma)}d\sigma
 \simeq \frac{\mu^4}{\nu^2} \int_{\sigma_c}^{\sigma}
         \frac{\sigma}{\sigma_d^4+\sigma^4} d\sigma
 \simeq \frac{\mu^4}{\nu^2}\frac{1}{2\sigma_d^2} 
        \arctan \lmk \frac{\sigma^2}{\sigma_d^2} \rmk
 \simeq 8.6.
\eeq 
Thus another inflation must take place to push the relevant scale to
the scale $k_0=0.002\,{\rm Mpc}^{-1}$.

Finally, from the constraint imposed on the amplitude of the tensor
perturbations by WMAPext+2dFGRS+Ly$\alpha$ \cite{Peiris:2003ff} we
find an upper bound on the energy scale of hybrid inflation as
\beq
  \mu < 1.4\times 10^{-2}, \label{tensor}
\eeq
corresponding to $H< 1.1\times 10^{-4}$.

\section{New inflation in supergravity}

\label{sec:new}

In order to push the scales with the desired spectral shape of density
fluctuations to cosmologically observable scales, we invoke new
inflation which follows hybrid inflation discussed above. New inflation
is also favorable in that it predicts low reheating temperature 
to avoid overproduction of gravitinos, contrary to
hybrid inflation. Though new inflation has a severe initial condition
problem in general, in our model, an appropriate initial condition for
 new inflation is
dynamically realized during hybrid inflation.
In this section, we briefly review new inflation induced by the
superpotential $W_N$ with the K\"ahler potential $K_N$ \cite{Izawa:1997df}.

The scalar potential derived from $W_N$ and  $K_N$ is given by
\bea
V_N[\Phi]&=&\frac{\exp\lmk |\Phi|^2+\frac{\kn}{4}|\Phi|^4\rmk}
{1+\kn|\Phi|^2} \nonumber \\
& &\times\lkk~ \left| \lmk 1+|\Phi|^2+\frac{\kn}{2}|\Phi|^4\rmk v^4
-\lmk 1+\frac{|\Phi|^2}{n+1}+\frac{\kn|\Phi|^4}{2(n+1)}\rmk
h\Phi^n \right|^2 \right. \nonumber \\ 
& &~~~\left.
-3\lmk 1+\kn|\Phi|^2 \rmk|\Phi|^2\left| v^2-\frac{h}{n+1}\Phi^n
\right|^2\rkk .
\eea
Then, the potential minimum is given by
\beq
 |\Phi|_{\min}\cong\lmk\frac{v^2}{h}\rmk^{\frac{1}{n}}
 ~~~{\rm and~~Im}\Phi^n_{\min}=0, 
\eeq
with a negative energy density
\beq
 V_N[\Phimin]\cong -3e^{K_N}|W_N[\Phimin]|^2\cong
 -3\lmk\frac{n}{n+1}\rmk^2v^4|\Phimin|^2.
\eeq
Assuming that this negative value is canceled by a positive
contribution due to supersymmetry breaking, $\Lambda_{\rm SUSY}^4$, we
can relate energy scale of this model with the gravitino mass
$m_{3/2}$ as
\beq
 m_{3/2}\cong \frac{n}{n+1}\lmk\frac{v^2}{h}\rmk^{\frac{1}{n}}v^2.
\eeq

Without loss of generality we may identify the real part of $\Phi$
with the inflaton $\phi\equiv\sqrt{2}{\rm Re}\Phi$.  The dynamics of
inflaton is governed by the lower-order potential
\beq
 V_N[\phi]\cong v^4-\frac{\kn}{2}v^4\phi^2-\frac{2h}{2^{n/2}}v^2\phi^n 
 +\frac{h^2}{2^n}\phi^{2n}. \label{Neffpote}
\eeq
Since the last term is negligible during inflation and the Hubble
parameter is dominated by the first term, $H=v^2/\sqrt{3}$, the
slow-roll equation of motion reads
\beq
 3H\dot{\phi}=-V'_N[\phi]\cong
-\kn v^4\phi-{2^{\frac{2-n}{2}}}{nhv^2}
\phi^{n-1},  \label{Neqm}
\eeq
and the slow-roll parameters are given by
\beq
 \epsilon\cong\frac{1}{2}\lmk\kn\phi+{2^{\frac{2-n}{2}}}{nh}
\frac{\phi^{n-1}}{v^2}\rmk^2,~~~
\eta=-\kn-{2^{\frac{2-n}{2}}}{n(n-1)h}\frac{\phi^{n-2}}{v^2},
\eeq
in this new inflation regime. Thus inflation is realized with $\kn \ll
1$ and ends at
\beq
  \phi=\sqrt{2}\lmk\frac{(1-\kn)v^2}{hn(n-1)}\rmk^{\frac{1}{n-2}}\equiv
  \phi_e,
\eeq
when $|\eta|$ becomes as large as unity.

Since the two terms on the right-hand side of Eq. (\ref{Neqm}) are
identical at
\beq
  \phi=\sqrt{2}\lmk\frac{\kn v^2}{hn}\rmk^{\frac{1}{n-2}}\equiv\phi_d,
\eeq
the number of $e$-folds of new inflation is estimated as
\beq
 N_N=-\int_{\phi_i}^{\phi_e}\frac{V_N[\phi]}{V'_N[\phi]}d\phi
 \cong\int_{\phi_i}^{\phi_d}\frac{d\phi}{\kn\phi} +
\int_{\phi_d}^{\phi_e}\frac{2^{\frac{n-2}{2}}v^2}{hn\phi^{n-1}}d\phi
=\frac{1}{\kn}\ln\frac{\phi_d}{\phi_i}+\frac{1-n\kn}{(n-2)\kn(1-\kn)},
\eeq
for $0<\kn<n^{-1}$. In the case that $\kn$ vanishes, we find
\beq
 N_N=\int_{\phi_i}^{\phi_e}
\frac{2^{\frac{n-2}{2}}v^2}{hn\phi^{n-1}}d\phi
=\frac{2^{\frac{n-2}{2}}v^2}{hn(n-2)}
\phi_i^{2-n}-\frac{n-1}{n-2}. \label{nnzero}
\eeq
Here $\phi_i$ is the initial value of $\phi$, whose determination
mechanism is discussed in the next section.

\section{Hybrid new inflation in supergravity with a chaotic initial 
condition}

\label{sec:hybridnew}

In this section, we will investigate the whole dynamics of our model
by considering the total superpotential (\ref{eq:totalsuper}) and
the K\"ahler potential (\ref{eq:totalK}), under the condition
\beq
  v^4 \ll \mu^4.  \label{new}
\eeq

Initially, the potential is dominated by the term $\nu^2 \sigma^4 /4$ so
that inflation starts with the chaotic scenario. As
inflation proceeds and the inflaton $\sigma$ falls to sub-Planckian
scale, the
false vacuum energy $\mu^4$ becomes dominant so it turns to the usual hybrid
inflation scenario.  In this epoch, the mass terms of the other fields
are given by
\bea
  V &\supset& \lmk \frac{\nu^2}{4}\sigma^4 + \mu^4 + v^4 \rmk \varphi^2
            + \lmk \frac{\nu^2}{4}\sigma^4 + \kx \mu^4 + v^4 \rmk |X|^2
            \non \\
         && + \lmk \kz \frac{\nu^2}{4}\sigma^4 + \mu^4 + v^4 
                  + 2 \nu^2 \sigma^2 \rmk |Z|^2 
            + \lmk \frac{\nu^2}{4}\sigma^4 + \mu^4 + v^4
                 + \frac{g^2}{2} \sigma^2 \rmk 
            ( |\Psi|^2 + |\overline{\Psi}|^2 ) \non \\
         && + \lmk \frac{\nu^2}{4}\sigma^4 + \mu^4 - \kn v^4 
                  + u^2 \rmk |\Phi|^2 \non \\
         && - \lambda \mu^2 ( \Psi\overline{\Psi} 
              + \Psi^{\ast}\overline{\Psi}^{\ast} )
            - \frac{\nu}{2} \sigma^2 u (\Phi + \Phi^{\ast}) \non \\
         && + \mu^2 v^2 ( X \Phi^{\ast} + X^{\ast} \Phi ) 
          + \frac{\nu}{2} \sigma^2 v^2 ( Z \Phi^{\ast} + Z^{\ast} \Phi ) 
          - \frac{\nu}{2} \sigma^2 \mu^2 ( X Z^{\ast} + X^{\ast} Z ) 
          \non \\
          &=&  \lmk \frac{\nu^2}{4}\sigma^4 + \mu^4 + v^4 \rmk \varphi^2
            + M_{-}^2 |\Psi_{+}|^2 + M_{+}^2 |\Psi_{-}|^2 \non \\
         && + \lmk \frac{\nu^2}{4}\sigma^4 + \kx \mu^4 + v^4 \rmk |X|^2
            + \lmk \kz \frac{\nu^2}{4}\sigma^4 + \mu^4 + v^4 
                  + 2 \nu^2 \sigma^2 \rmk |Z|^2 \non \\          
         && + \lmk \frac{\nu^2}{4}\sigma^4 + \mu^4 - \kn v^4 
                  + u^2 \rmk |\Phi|^2 
            - \frac{\nu}{2} \sigma^2 u (\Phi + \Phi^{\ast})
            \non \\
         && + \mu^2 v^2 ( X \Phi^{\ast} + X^{\ast} \Phi ) 
          + \frac{\nu}{2} \sigma^2 v^2 ( Z \Phi^{\ast} + Z^{\ast} \Phi ) 
          - \frac{\nu}{2} \sigma^2 \mu^2 ( X Z^{\ast} + X^{\ast} Z ). 
\eea
Clearly, the amplitudes of $\varphi$, $\Psi$, and $\overline{\Psi}$
rapidly vanish during hybrid inflation. When $\sigma$ becomes
equal to $\sigma_c$, $M_{-}^2$ becomes negative so that the phase
transition occurs and hybrid inflation terminates.  On the other hand,
due to the presence of the term proportional to $\Phi + \Phi^{\ast}$,
$\Phi$ deviates from the origin during inflation. Then, due to the cross
terms with $\Phi$, $X$ and $Z$ also deviate from the origin. The
field values at the potential minimum can be estimated as
\bea
  X_{\min} &\simeq& - \frac{4 u \mu^2 v^2}{\nu^3 \sigma^6}
                       \frac{1+\kz}{\kz}, \non \\
  Z_{\min} &\simeq& - \frac{2 u v^2}{\nu^2 \sigma^4}
                       \frac{1}{\kz}, \non \\
  \Phi_{\min} &\simeq& \frac{u}{\nu\sigma^2},
\eea
for $\nu^2 \sigma^4 / 4 \gg \mu^4$, and
\bea
  X_{\min} &\simeq& - \frac{u \nu \sigma^2 v^2}{2 \mu^6}
                       \frac{1}{\kx},  \non \\
  Z_{\min} &\simeq& - \frac{u \nu^2 \sigma^4 v^2}{4\mu^8}
                       \frac{1+\kx}{\kx}, \non \\
  \Phi_{\min} &\simeq& \frac{u \nu \sigma^2}{2 \mu^4},
\eea   
for $\nu^2 \sigma^4 / 4 \ll \mu^4$. Here we have assumed that $2\nu^2
\sigma^2 < \mu^4$ in the mass squared of $Z$, which yields the
constraint
\beq
  2 \sqrt{\lambda} \nu < g\mu.
  \label{eq:constraint1}
\eeq

It can be easily shown that the correction to the dynamics of hybrid
inflation due to the presence of the deviations from the origins can
be negligible if the following conditions are satisfied:
\beq
  u \ll \mu^2, \qquad \qquad \frac{\lambda u v^2}{g \mu^4} \ll 1. 
  \label{eq:constraint2}
\eeq
Then, all of the results derived in Sec. \ref{sec:hybrid} still hold
true.

Since the effective mass of the field $\phi = \sqrt{2}$ Re$\Phi$ is
larger than the Hubble parameter during hybrid inflation, $H_H$, the above
configuration is realized with the dispersion
\beq
 \langle(\phi-\phi_{\min})^2\rangle=\langle\chi^2\rangle
=\frac{3}{8\pi^2}\frac{H_H^4}{\mu^4}=\frac{\mu^4}{24\pi^2}
\eeq
due to quantum fluctuations \cite{BD}. The ratio of quantum
fluctuation to the expectation value should be less than unity,
\beq
 \frac{\sqrt{\langle(\phi-\phi_{\min})^2\rangle}}{|\phi_{\min}|}
=\frac{1}{4\sqrt{3}\pi} \frac{g^2 \mu^4}{u \nu \lambda } 
\ll 1,
\label{ratio}
\eeq
where we have used
\beq
  \phi_{\min} 
    \simeq \frac{u \nu \sigma_c^2}{\sqrt{2} \mu^4}
    \simeq \frac{\sqrt{2} u \nu \lambda}{g^2 \mu^2},
\eeq                          
corresponding to the value at the end of hybrid inflation,
$\sigma\simeq\sigma_c$.  Inserting the values $\nu \simeq 8.1\times
10^{-6}$ and $\mu^2/\lambda \simeq 2.2 \times 10^{-5}$ yields the
constraint
\beq
  \lambda h \ll 6.3 \times 10^4,
  \label{eq:constraint3}
\eeq
which is trivially satisfied. Thus the initial value of the inflaton
for new inflation is located off the origin with an appropriate
magnitude.

After hybrid inflation ends, $\phi$ oscillates and its amplitude
decreases with an extra factor $v/\mu$ by the time vacuum energy density
$v^4$ dominates the total energy density \cite{Kawasaki:1998vx}, 
so new inflation starts with
\beq
\phi_i = \frac{\sqrt{2} u \nu \lambda v}{g^2 \mu^3},
\eeq
and continues until
$\phi=\phi_e$ with the potential (\ref{Neffpote}).

Contrary to the hybrid inflation regime, we do not have significant
observational constraints on the new inflation regime. As stated in
the Introduction, generally speaking, the density fluctuations
generated during new inflation can become large because of the
smallness of the field value of the inflaton. However, since hybrid
inflation does not last so long, the comoving scale which leaves the
horizon at the beginning of new inflation is larger than 100 kpc so
that the density fluctuations on the corresponding scales should take
an appropriate value for structure formation, which is easy to realize
in this model, contrary to our previous model \cite{KYY}.  \if Thus,
we only impose on the parameters so that the density fluctuations
generated during new inflation are comparable with those produced
during hybrid inflation.  \fi

For definiteness, we consider the case with $n=4$ and $\kn=0$. Then, from
Eq. (\ref{nnzero}), the number of $e$-folds of new inflation reads
\beq
 N_N = \frac{v^2}{4h}\frac{1}{\phi_{i}^2} - \frac32.
\label{nnn}
\eeq
This should be around $40$ to push the comoving scale with appropriate
spectral shape to the appropriate physical length
scale,\footnote{Strictly speaking, extra $e$-folds $\ln(\mu/v)$ should
  be added in making a correspondence between comoving horizon scales
  during hybrid inflation and proper scales. This is because comoving
  scales that left the Hubble radius in the late stage of hybrid
  inflation reenter the horizon before the beginning of the new
  inflation. However, for simplicity, we set $N_N \simeq 40$.}  which
yields the relation
\beq
  \frac{u}{g^2 \sqrt{\lambda}} \simeq 
    \frac{6.9 \times 10^{-4}}{\sqrt{h}}.  \label{ugl}
  \label{eq:constraint4}
\eeq
Then the spectral index at the onset of new inflation, $\phi=\phi_i$,
reads
\beq
 n_s-1=-6\epsilon+2\eta\cong-12h^2\frac{\phi_i^6}{v^4}-12h\frac{\phi_i^2}{v^2}
\cong 2\eta \cong -0.07.
\eeq
On the other hand, the amplitude of curvature perturbation at the same
scale is given by
\beq
  \calr \simeq \frac{v^4}{4\sqrt{3}\pi h \phi_{i}^3}
        \simeq 98 \sqrt{h} v \simeq 10^{-4}.
  \label{eq:constraint5}
\eeq
Here we have required $\calr \simeq 10^{-4}$.
\if
 and set $h = 0.1$, which
yields $v \simeq 3.2 \times 10^{-6}$.

If we further set $\mu^2=\nu \simeq 6.9 \times 10^{-6}$, for example,
we find from (\ref{mln}),
\beq
 \lambda\simeq 0.33~~~{\rm and}~~~
  \mu \simeq 2.6 \times 10^{-3}.
\eeq
Inserting these values into the constraints (\ref{eq:constraint1}) and
(\ref{eq:constraint2}) using (\ref{ugl}), the parameter $g$ is constrained as
\beq
  3.0 \times 10^{-3} \ll g \ll 7.0 \times 10^{-2}.
\eeq
Setting, say, $g \simeq 9.0 \times 10^{-3}$ yields $u \simeq 1.1
\times 10^{-7}$.

Putting it all together, an example of viable sets of the model
parameters is given by $\lambda \simeq 0.33$, $h = 0.1$, $g \simeq 9.0
\times 10^{-3}$, $\mu \simeq 2.6 \times 10^{-3}$, $v \simeq 3.2 \times
10^{-6}$, $\nu \simeq 6.9 \times 10^{-6}$, and $u \simeq 1.1 \times
10^{-7}$, for which the results obtained by the WMAP can be completely
reproduced.  From the above values of the model parameters it is evident
that the scale of new inflation is much smaller than that of hybrid
inflation as promised. 

Of course, the above parameter set is just an
example. Other sets are permitted as long as they satisfy the
constraints 
\fi

Now we have listed all the constraints on the model parameters
(\ref{eq:constraint0}), (\ref{tensor}), (\ref{new}),
(\ref{eq:constraint1}), (\ref{eq:constraint2}),
(\ref{eq:constraint3}), (\ref{eq:constraint4}), and
(\ref{eq:constraint5}).  As listed in Table II, our model has seven
parameters after the spurion fields have acquired expectation values.
Among them, $\nu$ is fixed by the normalization of fluctuation
amplitude.  We have three more equalities, (\ref{eq:constraint0}),
(\ref{eq:constraint4}), and (\ref{eq:constraint5}), for the other six
undetermined parameters, $\lambda,~g,~h,~\mu,~v,$ and $u$.  Hence we
can express all these parameters in terms of three independent
variables. That is, if we fix the expectation values of the spurion
fields, we can describe all the parameters of the theory as functions
of $g'$, $\mu'$, and $v'$.  To do this let us take a simple view that
the expectation values of the spurions are all equal and normalize
them by $10^{-2}$, that is,
\beq
  \la \Xi\ra=\la\Pi\ra=\la\Sigma\ra\equiv 10^{-2}B,
\eeq
where $B$ is a parameter of order of unity.
Then by definition we find
\beq
g=10^{-2}g'B,~~ \mu= 10^{-2}\mu'B,~~\mbox{and}~~v=10^{-4}v'B.
\eeq
From Eqs. (\ref{eq:constraint0}), (\ref{eq:constraint4}), and
(\ref{eq:constraint5}), other parameters in Eq. (\ref{eq:totalsuper})
read
\bea
  \lambda&\simeq&4.4\mu'^2B^2,~~~~~~~~~~~
 h\simeq1.0\times 10^{-4}v'^{-2}B^{-4},\non\\
 u&=&10^{-4}u'B^2\simeq 1.4\times10^{-5}g'^2\mu'v'B^5, \label{other} \\
 \nu&=&10^{-4}\nu'B^2\simeq 8.4\times 10^{-6}. \non
\eea
Parameters in the original superpotential are given by
\bea
  \lambda'&\simeq&4.4\mu'^2B^2,~~~~~~~~~~~~~
 h'\simeq1.0\times 10^{-4}v'^{-2}B^{-4},\non\\
  u'&\simeq&0.14g'^2\mu'v'B^3,~~~~~\nu'\simeq 8.4\times 10^{-2}B^{-2}.
\eea

From Eqs. (\ref{tensor}), (\ref{new}), (\ref{eq:constraint1}),
(\ref{eq:constraint2}), and (\ref{eq:constraint3}), we find that $g'$,
$\mu'$, and $v'$ must satisfy
\bea
  \mu'~~ &<& 1.4B^{-1},~~~~~~~~~~
\mu'^{-1}v'~ \ll 10^2B^{-1},~~~~~~~~~~~g' > 0.36B^{-1}, \non \\ 
 g'^2\mu'^{-1}v' &\ll& 7.0B^{-3},~~~~~~~~g'\mu'^{-1}v'^3 \ll 1.6
  \times 10^2 B^{-6},\\
  \mu'v'^{-1} &\ll& 1.2\times 10^4 B. \non
\eea
If these inequalities are satisfied, which are in fact easily
satisfied with natural magnitudes of model parameters, our model is
viable.  As an example, let us take $B=1$, $g'=1$, $\mu'=0.3$, and
$v'=0.02$, which yields
\beq
\lambda=0.40,~~h=0.25,~~g=10^{-2},~~\mu=3.0\times 10^{-3},
~~v=2\times 10^{-6},~~u=8.6\times 10^{-8},
\eeq
with
\beq
\lambda'=0.40,~~h'=0.25,~~
u'=8.6\times 10^{-4},~~\mbox{and}~~\nu'=8.4\times 10^{-2}.
\eeq
In this case the gravitino mass takes an acceptable value,
$m_{3/2}=15$TeV.

\section{Discussion and Conclusion}

In this paper, we proposed an inflation model in supergravity, in which
hybrid inflation starts with a chaotic initial condition and new
inflation follows hybrid inflation.  In order to realize chaotic
inflation in supergravity, we introduced the Nambu-Goldstone-like shift
symmetry, which ensures the flatness beyond the gravitational scale
$M_{G}$. During hybrid inflation, adiabatic fluctuations with a running
spectral index with $\ns >1$ on a large scale and $\ns <1$ on a smaller
scale are generated, as inferred by the first-year data of WMAP.  The
initial condition of new inflation is also set dynamically during hybrid
inflation and we can acquire a sufficiently large number of $e$-folds
with the amplitude of fluctuation at its onset well under control.

The latter feature is especially important, because, if the amplitude
of fluctuation at the onset of new inflation, corresponding to $\sim
100$kpc today, turned out to be too large as in our previous model
\cite{KYY}, the Universe would be too clumpy on this scale with too
many dark-halo-like objects.  On the other hand, if we extrapolate the
best-fit spectrum obtained by the first-year WMAP data with a running
spectral index to smaller scales, galactic and smaller scale
fluctuations tend to be too small to realize timely galaxy formation.
In our model the amplitude of fluctuations generated during new
inflation can easily take appropriately larger values than that
obtained by simply extrapolating the large-scale power spectrum.  This
may be helpful for early star formation which is required for early
reionization \cite{Bennett:2003bz} and from the age estimate of
high-redshift quasars using the cosmological chemical clock \cite{jy}.

Finally we comment on the naturalness of our model.  Since our model
realizes three different inflationary scenarios in succession, its
Lagrangian is inevitably more complicated than that of a simple single
field model.  In particular, since the final form of the
superpotential contains seven parameters, whose order of magnitude
ranges from $10^{-7}$ to 1, and the energy scales of the latter two
inflation are strikingly different from each other, one may wonder if
it is too complicated and {\it ad hoc}.  If we return to a more
fundamental level of the theory with spurion fields, however, we find
that all the coupling parameters take values within a natural range of
$10^{-3}\sim 1$, and that it contains only a single energy scale,
namely, the expectation values of the spurion fields can be mutually
identified with $\CO(10^{-2})$, a typical unification scale.  At the
present level of our understanding of the fundamental theory, we may
only say that this scale is determined by cosmological observation,
namely, the amplitude of density fluctuations, as with simpler models
of single-field inflation, whose energy scale is usually determined by
the same observation.  Thus our model essentially has only one energy
scale and the huge difference of the energy scale between hybrid
inflation and subsequent new inflation is naturally realized by virtue
of the symmetries of the theory.

\acknowledgments{This work was partially supported by the JSPS
  Grant-in-Aid for Scientific Research No.\ 13640285 (J.Y.), and the
  JSPS (M.Y.). M.Y. is partially supported by the Department of Energy
  under Grant No. DEFG0291ER40688.}

\begin{table}[bht]
  \begin{center}
    \begin{tabular}{| c | c | c | c | c | c | c | c | c | c |}\hline
                & $X$ & $S$ & $\Psi$ & $\overline{\Psi}$ & $Z$ 
                   & $\Phi$ & $\Pi$ & $\Sigma$ & $\Xi$ \\
        \hline
        $Q_R$      & $\frac{n+2}{n+1}$ & $~0~$ & $\frac{n}{2(n+1)}$ 
                   & $\frac{n}{2(n+1)}$ & $-\frac{2}{n+1}$ 
                   & $\frac{2}{n+1}$ & $\frac{n}{2(n+1)}$
                   & $\frac{3n+4}{2(n+1)}$ & $\frac{n+2}{n+1}$ \\
        \hline 
        $G$        & $I$ & $I$ & $G$    & $\overline{G}$    & $I$    
                   & $I$    & $I$   & $I$      & $I$ \\
        \hline
        $Z_2$      & $+$ & $-$ & $+$    & $+$               & $+$      
                   & $+$    & $-$   & $-$       & $-$  \\ \hline
    \end{tabular}
    \caption{The symmetries and the charges of various superfields.} 
    \label{tab:charges}
  \end{center}
\end{table}

\begin{table}[bht]
\begin{center}
    \begin{tabular}{| c | c| c | p{70mm} |}\hline
Parameter\rule[-3.5mm]{0mm}{9.5mm} & Origin & \parbox{18mm}{Order of magnitude}  & Role  \\ \hline
$\mu$ & $\mu'\la \Pi\ra$ & $10^{-3}$ & energy scale of hybrid inflation
\rule{0mm}{6mm} \\
$v$   & $v'\la \Pi\ra^2$ & $10^{-6}$ & energy scale of new inflation \rule{0mm}{6mm}\\
$u$   & $u'\la \Pi\ra\la \Sigma\ra$ & $10^{-7}$ & interaction between
     hybrid inflaton and new inflaton 
     \\
$\nu$ & $\nu'\la \Xi\ra^2 $ &  $10^{-6}$ & square root of the self-coupling of chaotic inflaton
     \\
$\lambda$& $\lambda'$ & $0.1$ & tachyonic mass parameter at the end of  hybrid inflation \\
$g$& $g'\la \Xi\ra$  & $10^{-2}$ & mass of hybrid inflaton \rule{0mm}{6mm}\\
$h$ & $h'$  & $0.1$ & self-coupling of new inflaton \rule{0mm}{6mm}\\ \hline
\end{tabular}
   \caption{List of model parameters.  The expectation values of the
 spurion fields are taken to be 
mutually identical: $\la \Xi\ra=\la \Pi\ra=\la \Sigma\ra=10^{-2} B.$
}
   \label{tab:parameters}
\end{center}
\end{table}

\end{document}